\documentclass[aps,prd,nofootinbib,floatfix,twocolumn]{revtex4} 
\usepackage{epsfig}
\usepackage{graphicx}
\usepackage{dcolumn}

\begin{document}

\def\figurebox#1#2#3{%
  	\def\arg{#3}%
  	\ifx\arg\empty
  	{\hfill\vbox{\hsize#2\hrule\hbox to #2{\vrule\hfill\vbox to #1{\hsize#2\vfill}\vrule}\hrule}\hfill}%
  	\else
   	{\hfill\epsfbox{#3}\hfill}%
  	\fi}

\def\ifb{\rm fb^{-1}}
\def\upsln{ \Upsilon}
\newcommand{\jpsi}{J/ \psi}
\def\upsln{\Upsilon (1S)}
\def\jpsix{\jpsi+X}
\def\jpsiee{\jpsi\to e^+e^-}
\def\jpsimm{\jpsi\to \mu^+\mu^-}
\def\jpsill{\jpsi\to l^+l^-}
\def\upstojpsix{ \Upsilon (1S)\to \jpsi+X}
\def\upstochicx{ \Upsilon (1S)\to \chi_{cJ}+X}
\def\upstochic0{ \Upsilon (1S)\to \chi_{c0}+X}
\def\upstochic1{ \Upsilon (1S)\to \chi_{c1}+X}
\def\upstochic2{ \Upsilon (1S)\to \chi_{c2}+X}
\def\upstojpsiccg{ \Upsilon (1S)\to \jpsi c\bar{c}g+X}
\def\upstoqq{ \Upsilon (1S)\to q\bar{q}}
\def\upstomm{ \Upsilon (1S)\to\mu^+\mu^-}
\def\eetoqq{ \Upsilon (1S)\to q\bar{q}+X}
\def\eetomm{e^+e^-\to \mu^+\mu^-}
f
\def\eetojpsix{e^+ e^-\to\jpsi+X}
\def\ups4s{ \Upsilon (4S)}
\def\upsl4{ \Upsilon (4S)}
\def\psitwos{ \psi (2S) }

\title{Upsilon Decays at CLEO}

\author{Steven Blusk}

\affiliation{Syracuse University, Syracuse, NY 13244}


\begin{abstract}
Using data collected using the CLEO III detector, we present recent results 
on decays of the $\Upsilon(1S)-\Upsilon(3S)$ resonances. We report on
three analyses. They are: (1) improved measurements of the muonic 
branching fraction of the $\Upsilon(1S)-\Upsilon(3S)$, (2) precision 
measurements of $\Upsilon(2S)$ and $\Upsilon(3S)$ photonic transitions, 
and (3) new measurements of $\Upsilon(1S)$ decays to charmonium final
states.
\end{abstract}

\maketitle

\section{Introduction}
	Since their discovery in 1977, the $\Upsilon$
resonances have provided a unique laboratory for the study of strong interactions (QCD), 
while the $\Upsilon(4S)$ resonance provides a clean source of $B$ mesons which are
used to probe the weak interaction. 
The analyses reported on 
here make use of approximately 20$\times$10$^6$ $\Upsilon(1S)$, 9$\times$10$^6$ $\Upsilon(2S)$, and
$5\times 10^6$ $\Upsilon(3S)$ decays recorded using the CLEO III detector at CESR. The
analyses also utilize 0.2 $\ifb$, 0.4 $\ifb$ and 0.2 $\ifb$ of data taken just below 
the $\Upsilon(1S)$, $\Upsilon(2S)$, and $\Upsilon(3S)$ resonances ({\it ie., continuum}), 
respectively.

\section{Measurement of the Muonic Branching Fraction in $\Upsilon(1S)$, $\Upsilon(2S)$
and $\Upsilon(3S)$ Decays}
\label{sec:bmumu}
	
	The $\Upsilon$ resonances have provided a fertile testing ground for both 
lattice QCD and QCD-inspired potential models. Of particular interest are
the total widths of these resonances, $\Gamma$, which can be related to the muonic
branching fraction, ${\cal{B}}_{\mu\mu}$ via $\Gamma=\Gamma_{ee}/B_{\mu\mu}$, where 
the electronic decay width, $\Gamma_{ee}$, can be obtained from the integrated 
resonant cross-section for $e^+e^-\to hadrons$. $B_{\mu\mu}$ also enters
into the determination of other $\Upsilon$ branching fraction measurements, 
and therefore may present
a limiting systematic uncertainty. CLEO is now in a position to provide a much improved
measurement of ${\cal{B}}_{\mu\mu}$~\cite{bmumu}.

	The branching fraction ${\cal{B}}_{\mu\mu}$ can be extracted by measuring the
efficiency-corrected ratio: ${\tilde{\cal{B}}}=N_{\upsln\to\mu^+\mu^-}/N_{\upsln\to hadrons}$,
which is related to ${\cal{B}}_{\mu\mu}$ through ${\cal{B}}_{\mu\mu}={\tilde{\cal{B}}}/(1+{\tilde{\cal{B}}})$.
Dimuon events are selected by requiring exactly two nearly back-to-back charged tracks 
($N_{trk}$)
with energy $E_{\mu}$ in the range $0.7<E_{\mu}/E_{beam}<1.15$, one which must 
be detected in the muon system. 
Hadronic events are selected by requiring $N_{trk}\ge 3$, with
additional requirements on the energy in the calorimeter when $N_{trk}<5$. 
The number of dimuon (hadronic) events surviving all analysis
requirements are $344.9\times 10^3$ ($18.96\times10^6$), 
$119.6\times 10^3$ ($7.84\times10^6$), and $81\times 10^3$ ($4.64\times 10^6$) for 
$\Upsilon(1S)$, $\Upsilon(2S)$, and $\Upsilon(3S)$, respectively.
Event selection and analysis requirement efficiencies are determined from
simulation and are 65\% (96-98\%) for the dimuon (hadronic) event selection.
Backgrounds are estimated using simulation and are below 3\% for the 
dimuon samples and less than 1\% for the hadronic samples. The branching 
fractions are found to be: ${\cal{B}}(\Upsilon(1S)\to\mu^+\mu^-)=(2.49\pm0.02\pm0.07)\%$,
${\cal{B}}(\Upsilon(2S)\to\mu^+\mu^-)=(2.03\pm0.03\pm0.08)\%$ and
${\cal{B}}(\Upsilon(3S)\to\mu^+\mu^-)=(2.39\pm0.07\pm0.10)\%$,
where the first uncertainty is statistical and the second systematic. The
$\Upsilon(1S)$ measurement is consistent with previous measurements, but the
$\Upsilon(2S)$ and $\Upsilon(3S)$ are larger by 2-3 standard deviations.
These latter two measurements provide factor of 2-3 improvement in the 
precision to which these branching fractions are known while the $\Upsilon(1S)$
result is by itself competitive with the world average.

\section{Radiative Decays of the $\Upsilon(2S)$ and $\Upsilon(3S)$}

	Measurements of the radiative decays of the $\Upsilon$ resonances provide
clues into the nature of the strong force and provides for unique tests of potential
models and lattice QCD. For example, the electric dipole transition rate,
$\Gamma_{E1}={\cal{B}}(\Upsilon(nS\to\chi_{bJ}(n^{\prime}P))\Gamma(nS)$ 
can be determined experimentally, and can in turn be used to extract the
corresponding matrix element. The matrix elements and the mass splittings can
be used to learn about the contributing spin-orbit and tensor interactions~\cite{tomasz}.

	The events for this analysis~\cite{radups} use the same hadronic event selection as
in the ${\cal{B}}_{\mu\mu}$ analysis. Radiative transitions are observed through their 
peaks in the inclusive photon energy spectrum. Photon candidates are required to
be isolated from all charged particles and have a shower profile consistent with
an electromagnetic shower. The predominant source of background arises from 
$\pi^0\to\gamma\gamma$ decay. Application of a $\pi^0$ veto for low energy
photons does not improve the signal significance due to accidental
combinations where the signal photon and a random photon have an invariant
mass near the $\pi^0$ mass. Consequently, the $\pi^0$ veto is not applied to
the $\Upsilon(3S)\to\chi_{bJ}(2P)$ and $\Upsilon(2S)\to\chi_{bJ}(1P)$ analyses.
On the other hand, a $\pi^0$ veto is applied to the 
$\Upsilon(3S)\to\chi_{bJ}(1P)$ analysis, where the photon energies are
much higher and accidental combinatorial losses are much smaller.

	The inclusive photon spectrum for $\Upsilon(2S)$ data is shown in
Fig.~\ref{fig:y2s_1p}. The top figure shows the data (points) and overlayed 
are the fits to the data (solid line) and the background contribution (dashed line).
The lower plot shows the background subtracted peaks along with the fit (solid line)
and the contributions from the individual $\chi_{bJ}$ states (dashed line).
The background is determined using a combination of continuum and  $\Upsilon(1S)$ data. 
The analogous distributions for $\Upsilon(3S)$ data are shown in Fig.~\ref{fig:y3s_2p}.
In both cases, large signals are observed for each of the transitions. The
yields, efficiencies, branching fractions and photon energies are summarized
in Table~\ref{tab:upsrad}. The branching fraction measurements are by themselves about
a factor of two better than the world average values. Using the inclusive photon spectrum, we also
observe the rare transition, $\Upsilon(3S)\to\chi_{b0}(1P)$, which is found to have 
a branching fraction of $0.30\pm0.04\pm0.10)\%$. This is the first measurement of this
transition rate.

\begin{table*}
\caption{Summary of results on $E1$ transitions in $\Upsilon(2S\to\chi_{bJ}(1P)$ 
and $\Upsilon(3S\to\chi_{bJ}(2P)$. Where uncertainties are shown, the first 
is statistical and the second is systematic.}\label{tab:upsrad}
\begin{tabular}{|r|c|c|c|c|} 
\hline
Process &   $N_{\gamma}(\times10^3)$ &	Eff. (\%) &  ${\cal{B}}$ (\%)  	& $E_{\gamma}$ (MeV) \\
\hline
$\Upsilon(2S)\to\chi_{b0}(1P)$ & 198$\pm$6 & 57 & $3.75\pm0.12\pm0.47$ & $162.56\pm0.19\pm0.42$ \\
$\to\chi_{b1}(1P)$ & 407$\pm$7 & 63 & $6.93\pm0.12\pm0.41$ & $129.58\pm0.09\pm0.29$ \\
$\to\chi_{b2}(1P)$ & 410$\pm$6 & 61 & $7.24\pm0.11\pm0.40$ & $110.58\pm0.08\pm0.30$ \\
\hline
$\Upsilon(3S)\to\chi_{b0}(2P)$ & 225$\pm$7 &  57 & $6.77\pm0.20\pm0.65$ & $121.55\pm0.16\pm0.46$ \\
$\to\chi_{b1}(2P)$ & 537$\pm$7 &  63 & $14.54\pm0.18\pm0.73$ & $99.15\pm0.07\pm0.25$ \\
$\to\chi_{b2}(2P)$ & 568$\pm$6 &  61 & $15.79\pm0.17\pm0.73$ & $86.04\pm0.06\pm0.27$ \\
\hline
\end{tabular}
\end{table*}

\begin{figure}[hbt]
\epsfxsize160pt
\figurebox{120pt}{160pt}{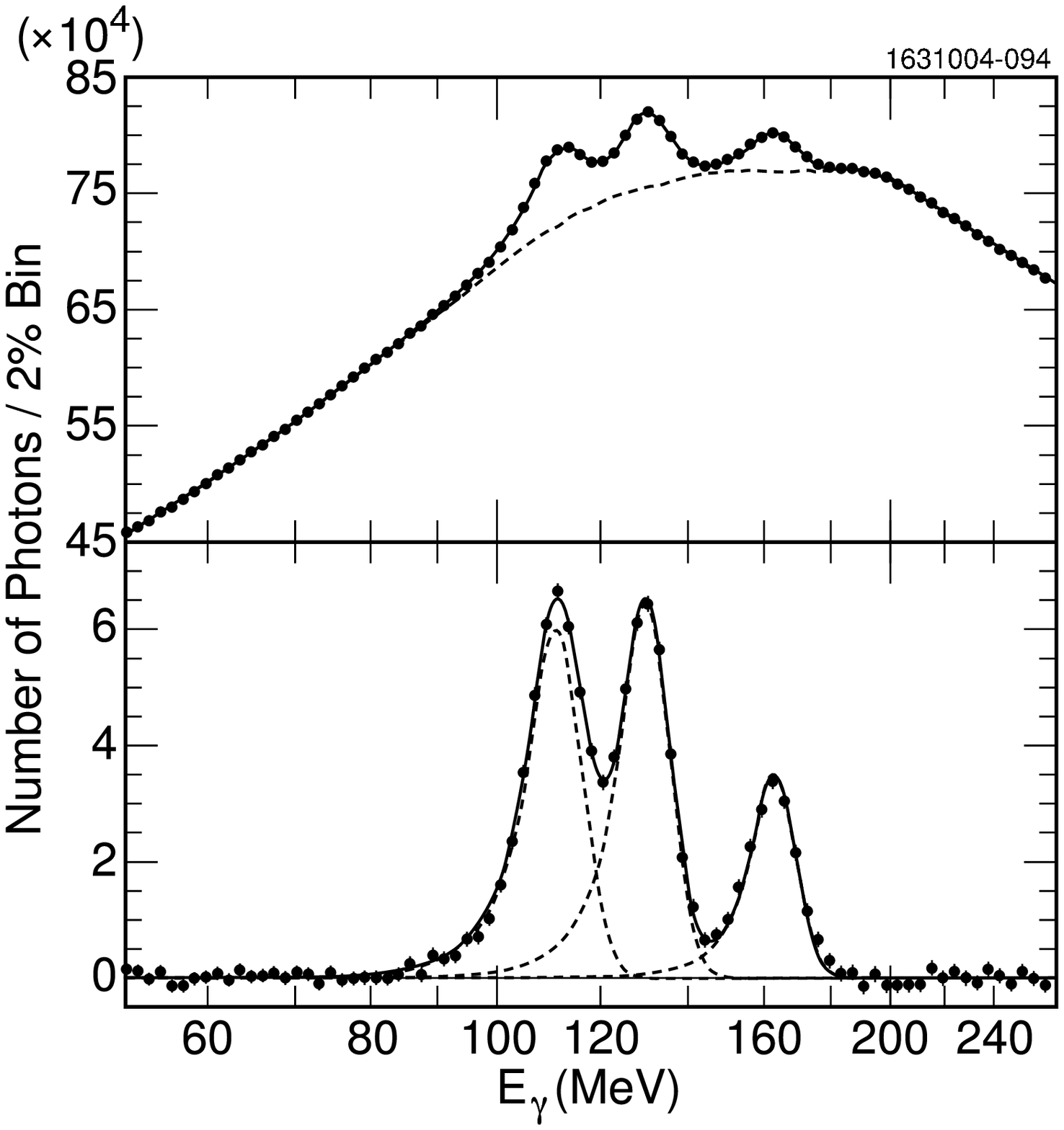}
\caption{Inclusive photon spectra in $\Upsilon(2S)$ data (top)
and the background subtracted spectrum (bottom). The solid line
is the combined fit and the dashed line show the contributions
from each of the $\chi_{cJ}$ states.}
\label{fig:y2s_1p}
\end{figure}
\vspace{-0.1in}

\begin{figure}[hbt]
\epsfxsize160pt
\figurebox{120pt}{160pt}{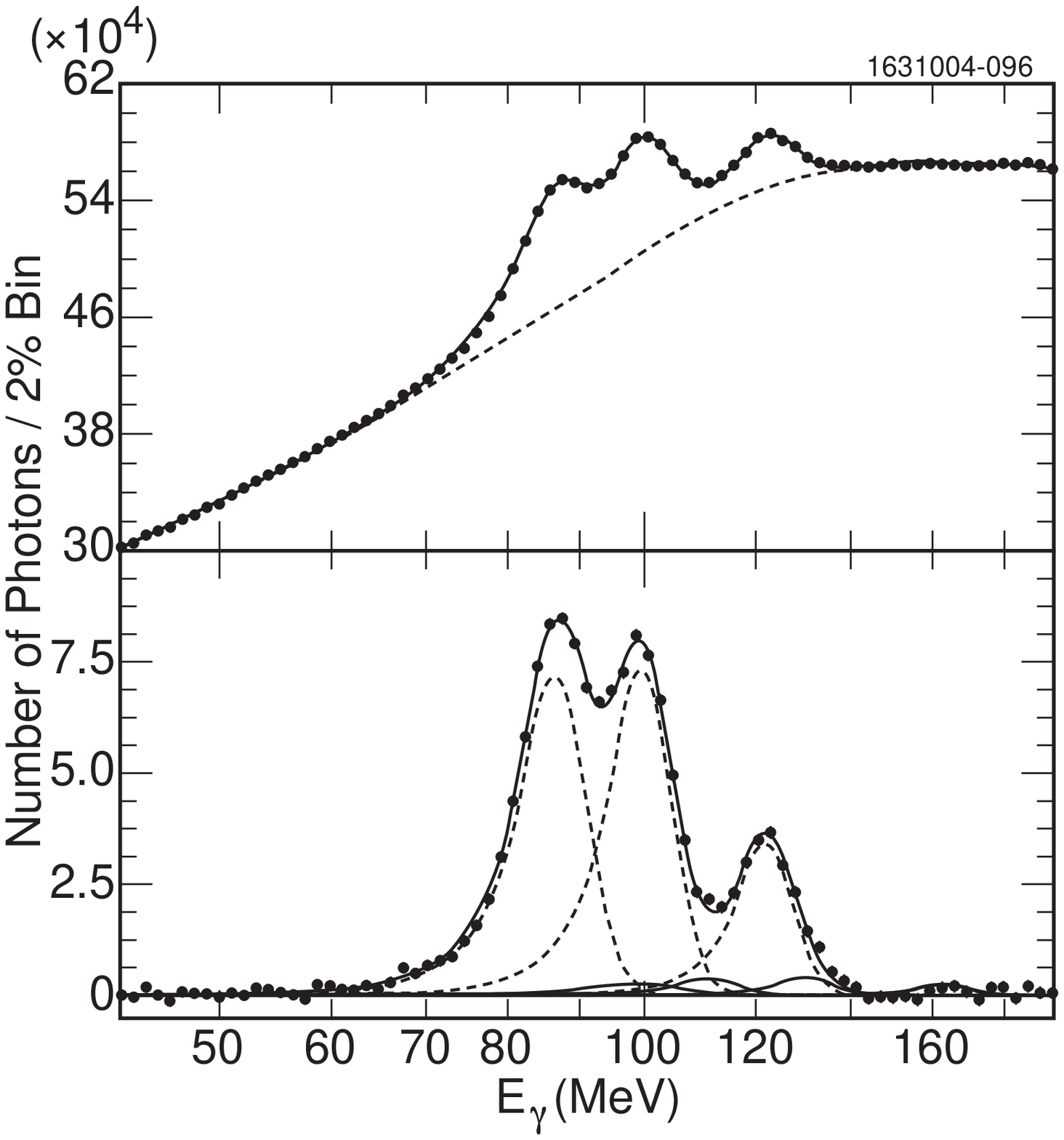}
\caption{Inclusive photon spectra in $\Upsilon(3S)$ data (top)
and the background subtracted spectrum (bottom). The solid line
is the combined fit and the dashed line show the contributions
from each of the $\chi_{cJ}$ states.}
\label{fig:y3s_2p}
\end{figure}
\vspace{-0.1in}

\section{New Measurements of Charmonium Production in $\Upsilon(1S)$ Decay}

	One of the longstanding problems in QCD has been to understand the
production mechanism for charmonium in a glue-rich environment. 
To account for the large excess of charmonium observed at the Tevtron, an
additional color-octet mechanism~\cite{braaten} was proposed whereby a 
$c\bar{c}$ pair is produced in a color-octet state, and then evolves 
non-perturbatively into a color-singlet. The matrix element
for the latter was determined as to account for the deficit. 
It has been suggested that the study of $\jpsi$ production
in three-gluon decays of $\Upsilon$ provides a unique opportunity to test color-octet
predictions. The most dramatic expectation is a peaking of the $\jpsi$ momentum
spectrum near the kinematic endpoint. A second mechanism which can also produce $\jpsi$
in $\Upsilon$ decay is the color-singlet process, $\upstojpsiccg$, which is inherently
soft and should yield two additional open charm particles.

	Events used in this analysis~\cite{blusk} are required to pass 
standard hadronic event selection. In order to suppress background from radiative
return ($e^+e^-\to\gamma\jpsi,\gamma\psi(2S)$), we reject events which have
$N_{trk}\le 4$ and either a detected photon with momentum greater than 3.75 GeV/c,
or an event momentum imbalance of larger than 2 GeV/c. Candidates $\jpsi$'s are 
reconstructed in the $\jpsimm$ and $\jpsiee$ modes. In addition to standard track 
selection criteria, muons are required to be detected 
in the muon system. Electrons are required to deposit energy in the calorimeter
which is consistent with its measured momentum. 
	
	The invariant mass distributions
for $\jpsimm$ and $\jpsiee$ candidates are shown in Fig.~\ref{fig:mass_1s}. The
continuum contribution ($e^+e^-\to\jpsi+X$) is estimated using data taken on
and below the $\Upsilon(4S)$. This contribution is added to the
expected $\Upsilon(1S)\to\gamma^*\to q\bar{q}\to\jpsi+X$ contribution, and the
total subtracted from the on-$\Upsilon(1S)$ data. The resulting 
scaled momentum distributions for $\jpsimm$ and $\jpsiee$ in $\Upsilon(1S)$
decays are shown in Fig.~\ref{fig:x_spec}. The data are compared 
to predictions of the color-octet (solid curve) and color-singlet (dashed curve).
The data are much softer than the color-octet predictions and moderately softer
than the color-singlet predictions. In the latter case, the effects of
fragmentation of the recoiling charm into $D$, $D^*$, $D^{**}$, $\Lambda_c$, etc
is not taken into account and is expected to further soften the spectrum. 
Integrating the spectrum, we measure a branching fraction
${\cal{B}}(\upstojpsix)=(6.4\pm0.4\pm0.6)\times10^{-4}$. This
measurement is consistent with predictions of both the color-octet and color-singlet
models which predict branching fractions of $6.2\times10^{-4}$ and 
$5.9\times10^{-4}$, respectively.

\begin{figure}[ht]
\epsfxsize180pt
\figurebox{120pt}{120pt}{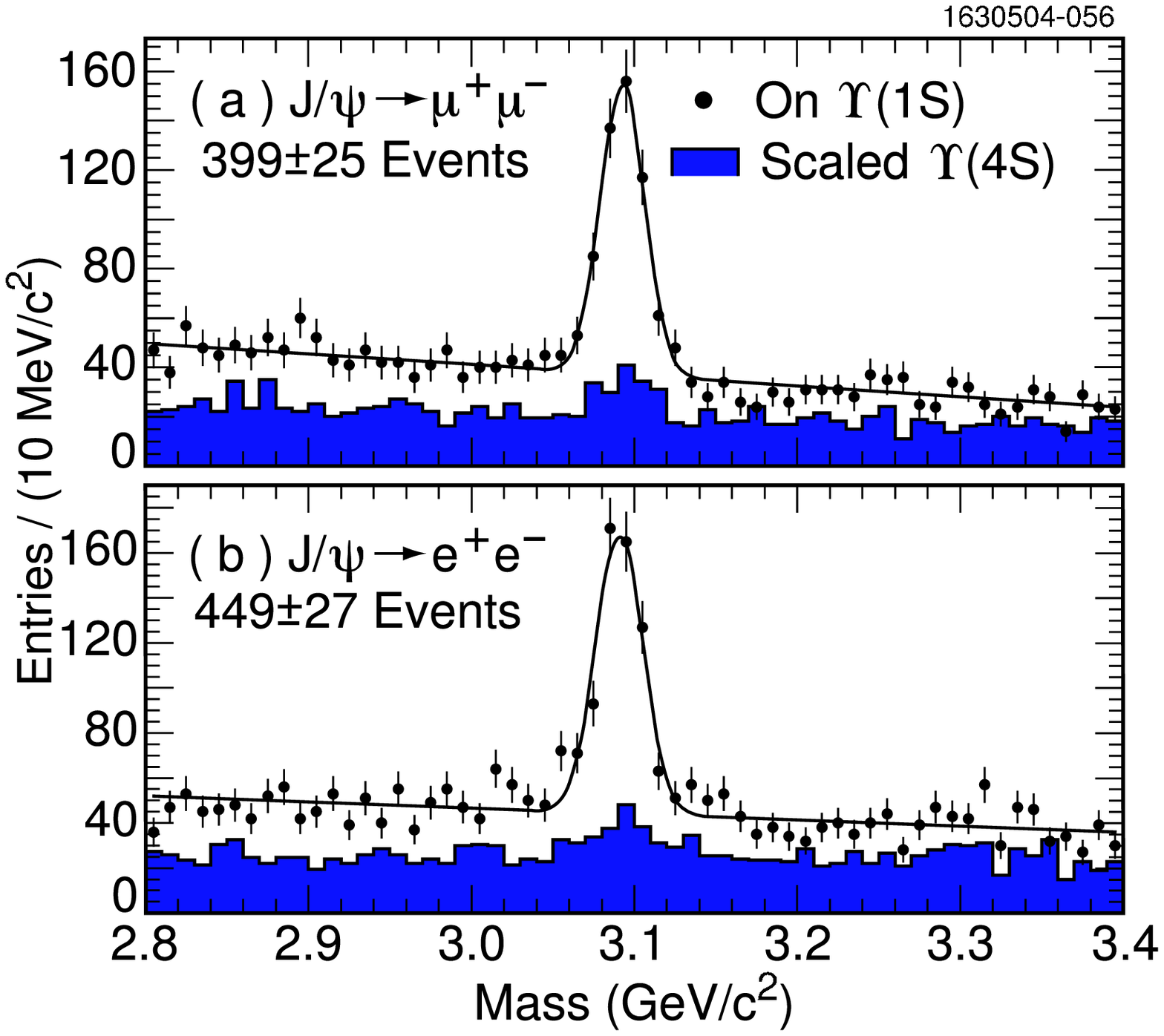}
\caption{Invariant mass distributions for $\mu^+\mu^-$ and $e^+e^-$ in
$\Upsilon(1S)$ data (points), and scaled $\Upsilon(4S)$ continuum data (shaded).}
\label{fig:mass_1s}
\end{figure}
\vspace{-0.1in}

\begin{figure}[ht]
\epsfxsize160pt
\figurebox{160pt}{120pt}{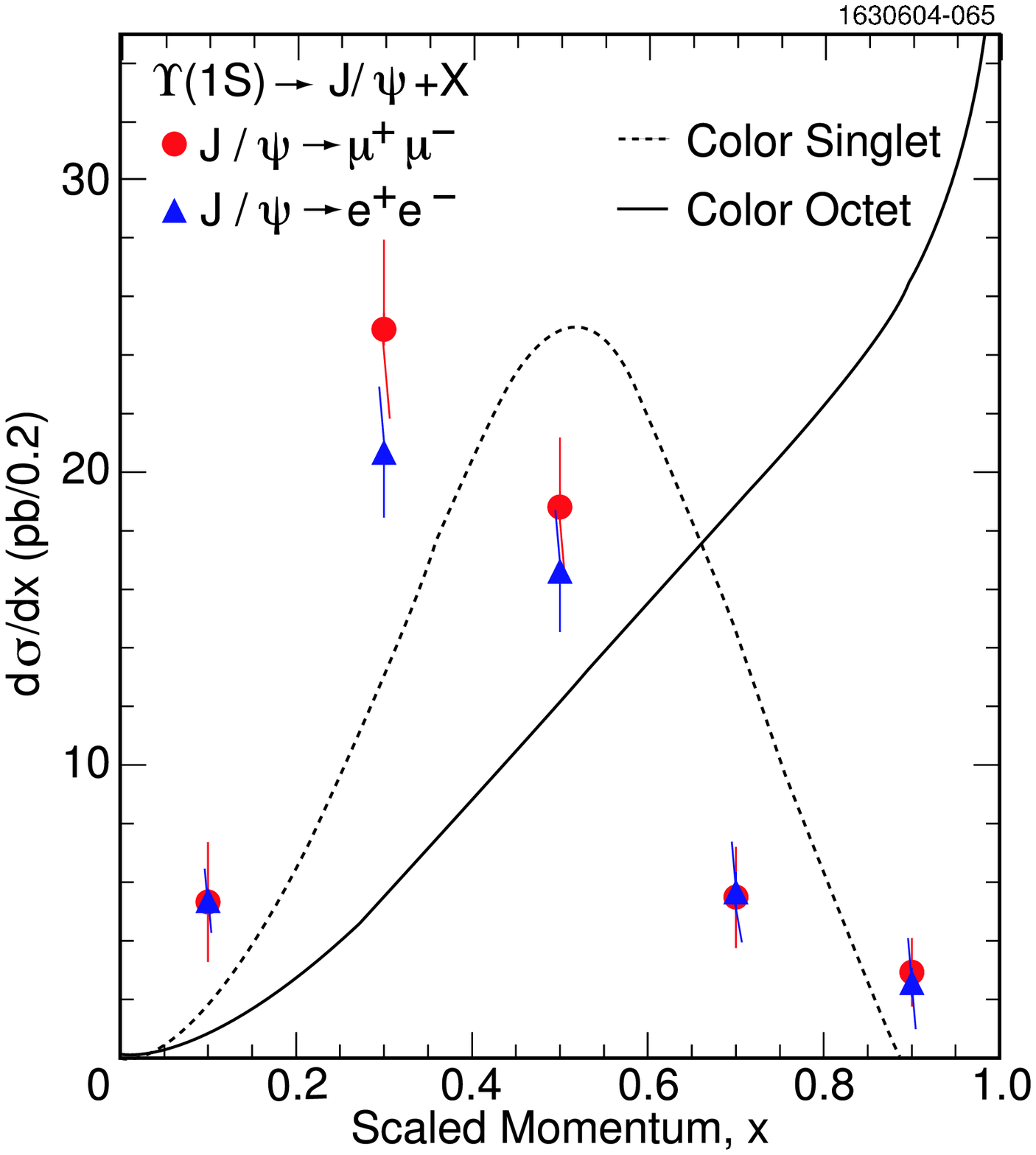}
\caption{Scaled momentum spectrum for $\upstojpsix$ in three-gluon decays
of the $\Upsilon(1S)$ for $\jpsimm$ (circles) and $\jpsiee$ (triangles). The
curves show the predictions of the color-octet and color-singlet models.}
\label{fig:x_spec}
\end{figure}
\vspace{-0.1in}

	We have also made a first observation of the decay, 
$\Upsilon(1S)\to\psi(2S)+X$ (see Fig.~\ref{fig:psi2s}) and found evidence 
for $\Upsilon(2S)\to\chi_{cJ}+X$.
The branching fractions are measured relative to $\upstojpsix$,
and are found to be:
${{\cal{B}}(\upsln\to\psitwos+X)\over {\cal{B}}(\upstojpsix )}  = 0.41\pm0.11\pm0.08$,
${{\cal{B}}(\upsln\to\chi_{c1}+X)\over {\cal{B}}(\upstojpsix)} = 0.35\pm0.08\pm0.06$, 
${{\cal{B}}(\upsln\to\chi_{c2}+X)\over {\cal{B}}(\upstojpsix)} = 0.52\pm0.12\pm0.09$, and
${{\cal{B}}(\upsln\to\chi_{c0}+X)\over {\cal{B}}(\upstojpsix)} < 7.4$ at 
90\% confidence level. 

	The feed-down contributions to $\jpsi$ are about a factor of two
higher than expected. These measurements have the potential to shed some 
additional light on the production mechanisms of charmonium in glue-rich
environments.

	I gratefully acknowledge the effort of the CESR staff in providing 
us with excellent luminosity and running conditions. I also thank
the National Science Foundation and the U.S. Dept. of Energy for their 
support of this research.

\begin{figure}[ht]
\epsfxsize160pt
\figurebox{120pt}{120pt}{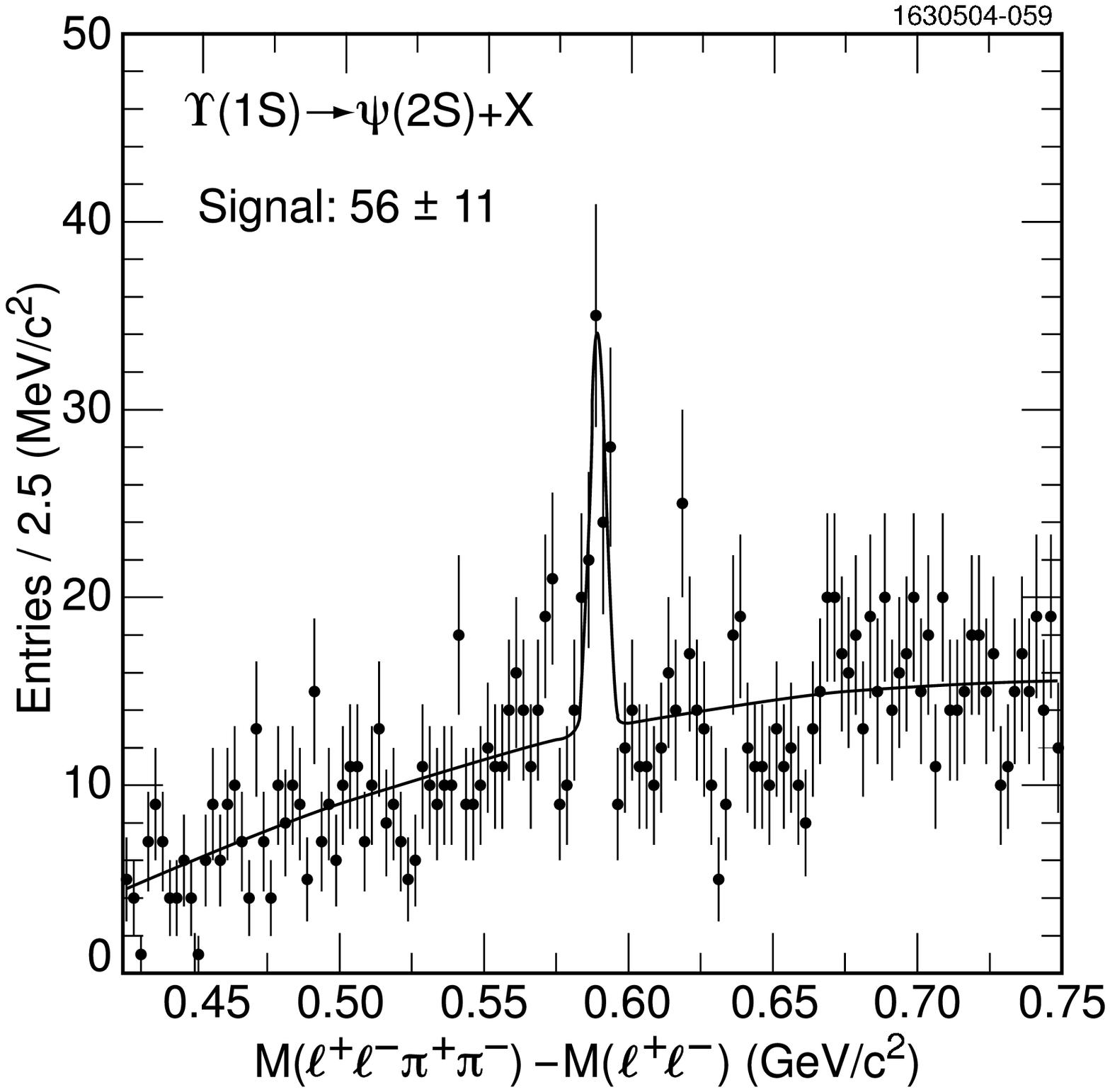}
\caption{Scaled momentum spectrum for $\upstojpsix$ in three-gluon decays
of the $\Upsilon(1S)$ for $\jpsimm$ (circles) and $\jpsiee$ (triangles). The
curves show the predictions of the color-octet and color-singlet models.}
\label{fig:psi2s}
\end{figure}
\vspace{-0.2in}


\end{document}